\documentclass[12pt]{iopart}
\usepackage{iopams}
\usepackage{epsf}

\usepackage{graphicx}
\usepackage{dcolumn}
\usepackage{bbm}
\usepackage{multirow}

\begin{document}

\title{Dynamics of a map with power-law tail}

\author{V. Botella-Soler\dag\ , J.A. Oteo\ddag\ and J.
Ros\dag\ \footnote[3]{To whom correspondence should be addressed} }

\address{\dag Departament de F\'{\i}sica Te\`{o}rica  and Instituto de
F\'{\i}sica Corpuscular, Universitat de Val\`{e}ncia,
46100-Burjassot, Val\`{e}ncia, Spain}

\address{\ddag\ Departament de F\'{\i}sica Te\`{o}rica, Universitat de
Val\`{e}ncia,  46100-Burjassot, Val\`{e}ncia, Spain}

\ead{vicente.botella@uv.es, oteo@uv.es, rosj@uv.es}

\begin{abstract}
{We analyze a one-dimensional piecewise continuous discrete model
proposed originally in studies on population ecology. The map is
composed of a linear part and a power-law decreasing piece, and has
three parameters. The system presents both regular and chaotic
behavior. We study numerically and, in part, analytically different
bifurcation structures. Particularly interesting is the description
of the abrupt transition order-to-chaos mediated by an attractor
made of an infinite number of limit cycles with only a finite number
of different periods. It is shown that the power-law piece in the
map is at the origin of this type of bifurcation. The system
exhibits interior crises and crisis-induced intermittency.}
\end{abstract}






\maketitle

\section{Introduction\label{sec:intro}}

Work on population ecology carried out in the 1970's certainly
helped chaos to move center stage as a new interdisciplinary subject
\cite{May2}. One-dimensional discrete-time models are technically
among the simplest ones to consider and interpret. They provide an
appropriate description of species with non-overlapping generations.
The basic example for the evolution of a population density $X_n \in
\mathbbm R$ at generation $n$ is the linear law $X_{n+1}=r X_{n}$.
Here $r$ represents the growth rate or fecundity, assumed constant.
It is, however, too schematic allowing only extinction ($r<1$),
equilibrium ($r=1$) or infinite growth ($r>1$). It was soon
recognized that more realistic models should be nonlinear:
\begin{equation}\label{eq:nonlinear}
X_{n+1}=X_n F(X_n).
\end{equation}
Written in this form, $F$ is the dimensionless non-constant fitness
function of the population or per-capita growth rate.
A sage choice for it should capture
the essential features of the system. The crucial point is that
once nonlinearity is let in, a huge variety of new phenomena may
appear as is now universally recognized. Thus, discrete time
population ecology is in close contact with discrete dynamical
system theory. For
instance, the choice \cite{mayetal,hassell}
\begin{equation}\label{eq:gradwell}
F(X_n)=r(X_n/K)^{-b}, \; b>0,
\end{equation}
renders easy the numerical determination of the parameter values
from experimental population data by a linear fit to $\log X_{n+1}$
versus $\log X_n$, which constitutes certainly a salient advantage.
In (\ref{eq:gradwell}) the presence of parameter $K$, the
conventional carrying capacity, ensures the dimensionless character
of $F$. A slight variant of it reads \cite{hassell,Varley}
\begin{equation}\label{eq:hassell}
F(X_n)=\left\{
\begin{tabular}{ll}
      $r,$ & $X_n\le C$ \\
      $r(X_n/K)^{-b},$ & $X_n>C$
     \end{tabular}
     \right.
 \nonumber
\end{equation}
where $C$ is a threshold population density and the fitness
parameter $b>0$ is dimensionless.

For the purposes of the present paper we emphasize  that these and
similar models can also be used as mathematical instances of
dynamical systems to illustrate different features when the ranges
of parameters and time variable are enlarged beyond those realistic
in population dynamics. In this spirit,
we analyze here the one-dimensional discrete model associated with
equation (\ref{eq:hassell}) which will be referred to as VGH
(Varley-Gradwell-Hassell) \cite{May1}.
In \cite{May3} it is briefly explained
that the system is chaotic for $b>2$ and regular for $b<2$, pointing
out that in the  order-to-chaos transition no cascade of period
doubling emerges. In the present paper we analyze numerically and
analytically how such a transition takes place which, to the best of
our knowledge, has not been studied in detail in the mathematics or
ecology literature.

The system function is continuous for $C=K$ and discontinuous
otherwise. Piecewise continuous maps are, by far, less analyzed than
continuous ones in the literature. Also, we address our attention to
the role of the threshold parameter $C$, to which no much attention
seems to have been given.

The paper is organized as follows. In Section \ref{sec:Descript} we
describe the features of the VGH map, present some traits of the
associated dynamical system, and briefly introduce the pertinent
equations to compute the Lyapunov exponent. In Section \ref{sec:b}
the transition order-to-chaos is studied in detail. Some exact
results are presented for $b=2$ and an explanation for the observed
bifurcation is propounded. In Section \ref{sec:xi} the effect of
varying $C$ and $r$ is analyzed. Finally, Section \ref{sec:fin}
contains our conclusions. For the sake of completeness, two
technical Appendices are included. In particular,
\ref{app:bookkeeping} contains a thorough analytical study of the
VGH map in the critical case $b=2$.

\section{Description of the map and its dynamics\label{sec:Descript}}

\subsection{Alternative formulations}
We write the VGH map, equation (\ref{eq:hassell}), in the form
\begin{equation} \label{eq:Iterate}
x_{n+1}=f(x_n;b,c,r)
\end{equation}
with
\begin{equation} \label{eq:Varley1}
f(x;b,c,r)=\left\{
\begin{tabular}{ll}
      $rx,$ & $x\le c$ \\
      $rx^{1-b},$ & $x>c$
     \end{tabular}
     \right.
 \nonumber
\end{equation}
In going from (\ref{eq:nonlinear}) and (\ref{eq:hassell}) to
(\ref{eq:Iterate}) and (\ref{eq:Varley1}) we have used the carrying
capacity $K$ as our natural unit to measure population densities and
correspondingly introduced the dimensionless variable $x_n=X_n/K$,
and parameter $c=C/K$.

The function $f$ in (\ref{eq:Varley1}) is defined in
${\mathbbm R}^4$ although, being a population variable,
we take $x\in
[0,\infty)$. In the three dimensional parameter space $(b,c,r) $ we
consider only the region $b>1$ to ensure decrease of $f$ with
increasing $x$; $c,r>0$ for compatibility with the range for $x$.

\begin{figure}[H]
\includegraphics[scale=1.0]{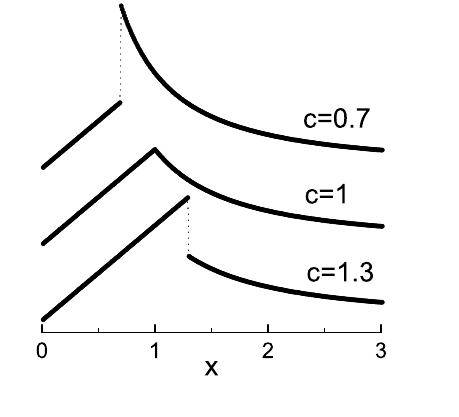}
\caption{\label{fig:V_plot} Shape of the map (\ref{eq:Varley1}) for
$b=2.5$, $r=2$ and three different values $c=0.7,1,1.3$. The curves
have been vertically shifted for the sake of clarity.}
\end{figure}

According to this map the population density in a given generation
is a linear function of that in the previous one as far as  it does
not exceed a critical value $c$. For greater values the population
follows a nonlinear power-law. Figure \ref{fig:V_plot} illustrates
the different possibilities and we expect, therefore, differences in
the response of the system according to whether the value of $c$ is
below or above unity.

For some purposes, we have found it useful to express the VGH map in
terms of the new variable $z$ and the new parameter $\xi$
\begin{equation}\label{eq:xi}
z\equiv 2\log(x)/\log(r), \quad \xi \equiv 2\log(c)/\log(r)
\end{equation}
with $x\ne 0$ and $r>1$.
The numeric factor two in (\ref{eq:xi}) has
been introduced for convenience. The inverse transformation reads
\begin{equation}
x=r^{z/2}, \quad c=r^{\xi/2}.
\end{equation}
This leads from (\ref{eq:Varley1}) to
\begin{equation} \label{eq:zlog}
z_{n+1}=\left\{
\begin{tabular}{ll}
      $z_n + 2,$ & $z_n\le \xi$ \\
      $(1-b)z_n +2,$ & $z_n>\xi$
     \end{tabular}
     \right.
 \nonumber
\end{equation}
now with phase space $(-\infty,\infty)$. This representation of the
map has just two effective parameters.  From a mathematical point of
view this transformed version is piecewise linear, whereas
(\ref{eq:Varley1}) defines a piecewise continuous nonlinear system.
Linearization is a standard procedure in the study of dynamical
systems. It usually follows from first order approximations. Here,
however, the linearization is exact. In \ref{app:lin} we collect
some results obtained from this version of the dynamical system. It
is worth mentioning that dynamical systems of the same type can be
found in applications in electronics, robotics and mechanical
systems with impacts \cite{power, Hogan}.

\subsection{VGH dynamical system}
We present here some of the most apparent features of the dynamical
system originated by the iteration of the VGH map.

First we can dispose of the case $r<1$. $x=0$ is then always an
stable fixed point. When $c<r^{1/b}$ there is a second fixed point
at $x=r^{1/b}$ which is stable if $1<b<2$ and which attracts any
initial condition in the interval $(c,x^*)$ with
$x^*=(c/r)^{1/(1-b)}$. For $b=2$, that interval consists of period-2
points, with the exception of the fixed point $\sqrt r$. Values
$x\le  c$ and  $x> x^*$ go to $x=0$. Observe that here $r<1$ does
not necessarily imply extinction.This is due to the interplay
between the different parameters.

For $r=1$ any initial condition in $[0,c]$ remains fixed. If $c<1$
then $x=1$ is also fixed. It is stable for $1<b<2$ attracting any
initial condition in the interval $(c,x^*)$ with $x^*=c^{1/(1-b)}$.
If $b=2$, the interior points of that interval, except $x=1$, are
paired in 2-cycles. Points to the right of $x^*$ are eventually
fixed, reaching its limiting fixed value in two steps.

For $r>1$, $x=0$
is an unstable fixed point, independently of $c$. If $c<r^{1/b}$, also $x=r^{1/b}$
is fixed, although it is stable only for $1<b<2$.

In the rest of the paper we consider only $r>1$ which, as we shall
see, generates a much richer behavior.

\subsection{Lyapunov exponent \label{sec:Lyap}}

The Lyapunov exponent is a measure of the rate at which two
initially close trajectories move away. For the one-dimensional
discrete map $x_{n+1}=f(x_n)$ the computation of the Lyapunov
exponent $\lambda$ admits an analytical formula (see for instance
\cite{Hilborn,Strogatz,Collet,Ott})
\begin{equation}\label{eq:L1}
\lambda=\lim_{n\to \infty}\left\{ \frac{1}{n}\sum_{k=0}^{n-1} \ln
|f'(x_k)|\right\}
\end{equation}
which, in the case (\ref{eq:Varley1}), leads to
\begin{equation}\label{eq:L2}
\lambda=\ln r+\lim_{n\to \infty}\left\{ \frac{1}{n}\sum_{k=0}^{n-1}
[\ln|1-b|-b\ln|x_k|]\theta(x_k-c)\right\} ,
\end{equation}
whereas (\ref{eq:zlog}) yields
\begin{equation}\label{eq:L2z}
\lambda=\ln|1-b|\lim_{n\to \infty}\left\{
\frac{1}{n}\sum_{k=0}^{n-1} \theta(z_k-\xi)\right\} .
\end{equation}
Both formulas are given in terms of Heaviside $\theta$ function
which selects the iterations that visit $x_n>c$ or $z_n>\xi$
respectively.

The mathematical equivalence of the two expressions for $\lambda$ in
(\ref{eq:L2}) and (\ref{eq:L2z}) yields the following result for the
statistics of points $x_k>c$ in the attractor
\begin{equation}
\lim_{n\to \infty}\left[ \frac{1}{n}\sum_{k=0}^{n-1} \ln
(x_k)\theta(x_k-c)\right] =\frac{\ln r}{b}
\end{equation}
which tells that the geometric mean of all the points $x_k>c$ in the
trajectory equals the value $r^{1/b}$. A simpler version of
(\ref{eq:L2}) reads now
\begin{equation}\label{eq:L2simple}
\lambda=\ln|1-b|\lim_{n\to \infty}\left\{
\frac{1}{n}\sum_{k=0}^{n-1} \theta(x_k-c)  \right\}.
\end{equation}

Equation (\ref{eq:L1}) makes use of the derivative $f'$ which in the
present case is not defined at $x=c$ or $z=\xi$. As it is an
isolated point, it does not amount any numerical difficulty.

In practice, one has to take care of discarding transients in order
to allow the orbit to enter the attractor. In particular, transients
are very long for chaotic trajectories when $b\approx 2$. Besides,
it is convenient to average $\lambda$ over a number of initial seeds.

The interpretation of the Lyapunov exponent in this system is
particularly straightforward: $\lambda$ is ruled by the proportion
of exterior points in the trajectories, namely those with $x_n>c$ or
$z_n>\xi$.

Since the summation in (\ref{eq:L2}) and (\ref{eq:L2z}) is over a
large enough number of points on the trajectory, numerical accuracy
is a relevant issue. To this end, and following the analysis in
\cite{OR}, we have used the multi-precision Fortran package MPFUN
\cite{Bailey, Bailey1} which allows one to efficiently compute with
very high number of digits. We concluded that double precision gives
here enough computational guaranties as regards (\ref{eq:L2}) and
(\ref{eq:L2z}). By contrast, it is not the case for the computation
of trajectories of (\ref{eq:zlog}) with some integer values $b>2$,
an issue that we deal with in Section \ref{sec:b-integer}.

\section{Dependence on $b$:  order-to-chaos transition \label{sec:b}}
As already stated in \cite{May3}, the chaotic regime corresponds to
$b>2$, independently of the values of $c$ and $r$, without stable
regularity windows embedded. This is clear from the positive
character of $\lambda$ in (\ref{eq:L2simple}) when $b>2$.

In the limit $b\to 2^+$ the Lyapunov exponent becomes very small due
to the term $\ln|1-b|$, since the proportion of exterior points is
finite. Following results on nonextensive thermodynamics it may be
pertinent to wonder whether a power law $n^\gamma$ is a more
appropriate ansatz than the exponential $\exp(\lambda n)$ to measure
the divergence of initially close trajectories \cite{Costa,
Tsallis}. We have carried out a statistics of the largest
separations, in this limit, as a function of $n$ and the results
point out towards a clear exponential divergence.

\begin{figure}[H]
\includegraphics[scale=1.3]{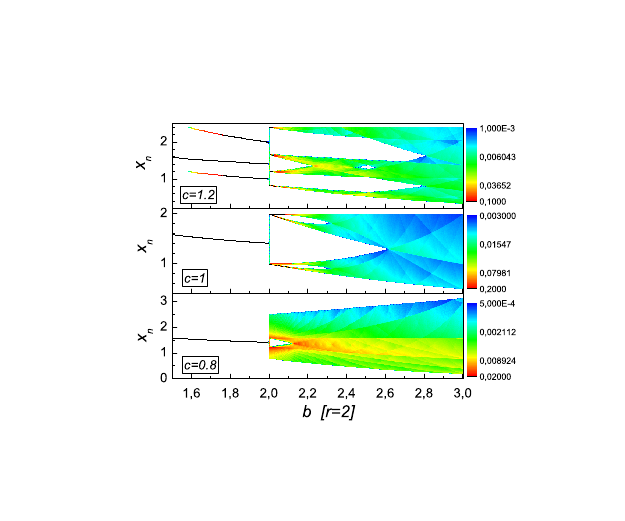}
\caption{\label{fig:V_r2_bif-b}   (Color online) Bifurcation
diagrams as a function of $b$,  for $r=2$ and three different values
of $c$. The color scale is logarithmic and it stands for the
frequency the point is visited with. This holds for the rest of
color figures, unless otherwise stated.}
\end{figure}

\begin{figure}[H]
\includegraphics[scale=1.2]{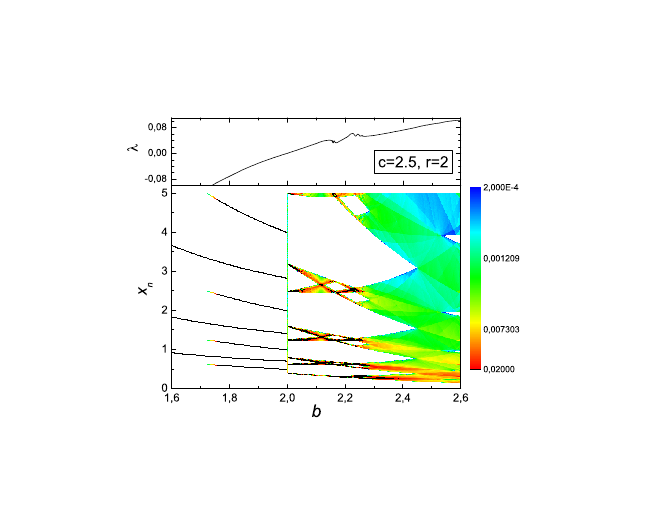}
\caption{\label{fig:V_C25_r2_xLy}   (Color online) Bifurcation
diagram as a function of $b$,  for $r=2,\; c=2.5$. The corresponding
Lyapunov exponent is in the upper panel. }
\end{figure}

The behaviour of the system with respect to $b$ is illustrated by
the bifurcation diagrams in Figure \ref{fig:V_r2_bif-b}. The three
panels correspond to $c=1.2,1,0.8$, from top to bottom,
respectively, with the fixed value $r=2$. In Figure
\ref{fig:V_C25_r2_xLy} we plot the bifurcation diagram for $c=2.5$
and $r=2$, as well as the Lyapunov exponent $\lambda$ (upper panel).
Notice the irregularities in the curve around the value $b=2.2$
where the bifurcation diagram exhibits a number of band merging
crises \cite{Ott, Grebogi-Ott}.

The aforementioned sudden transition from a regular to a chaotic
system, with no period doubling, is apparent. As the negative
sign of the computed Lyapunov exponent witnesses, $b<2$ means
regular behavior: every initial seed ends up in a periodic orbit.

For fixed $b<2$, the attractor
consists of coexisting limit cycles. Its cardinal is a piecewise
constant function of $c$. So, there exist some values $c=c_n$ ($n\ge 1$
integer) that punctuate the sequence of the number of points that
make up the attractor. The cardinal is given by the heuristic
formula $ 2(n+1)/[3-(-1)^n]$, $(n\ge 1)$, whose first few terms read
\begin{equation}\label{eq:seq}
1,3,2,5,3,7,4,9,5,11,6,13,7,15,8,17,9,\ldots
\end{equation}
They can be interpreted as forming two intermingled sequences whose
terms increase by one and two units respectively.

\begin{figure}[H]
\includegraphics[scale=0.55]{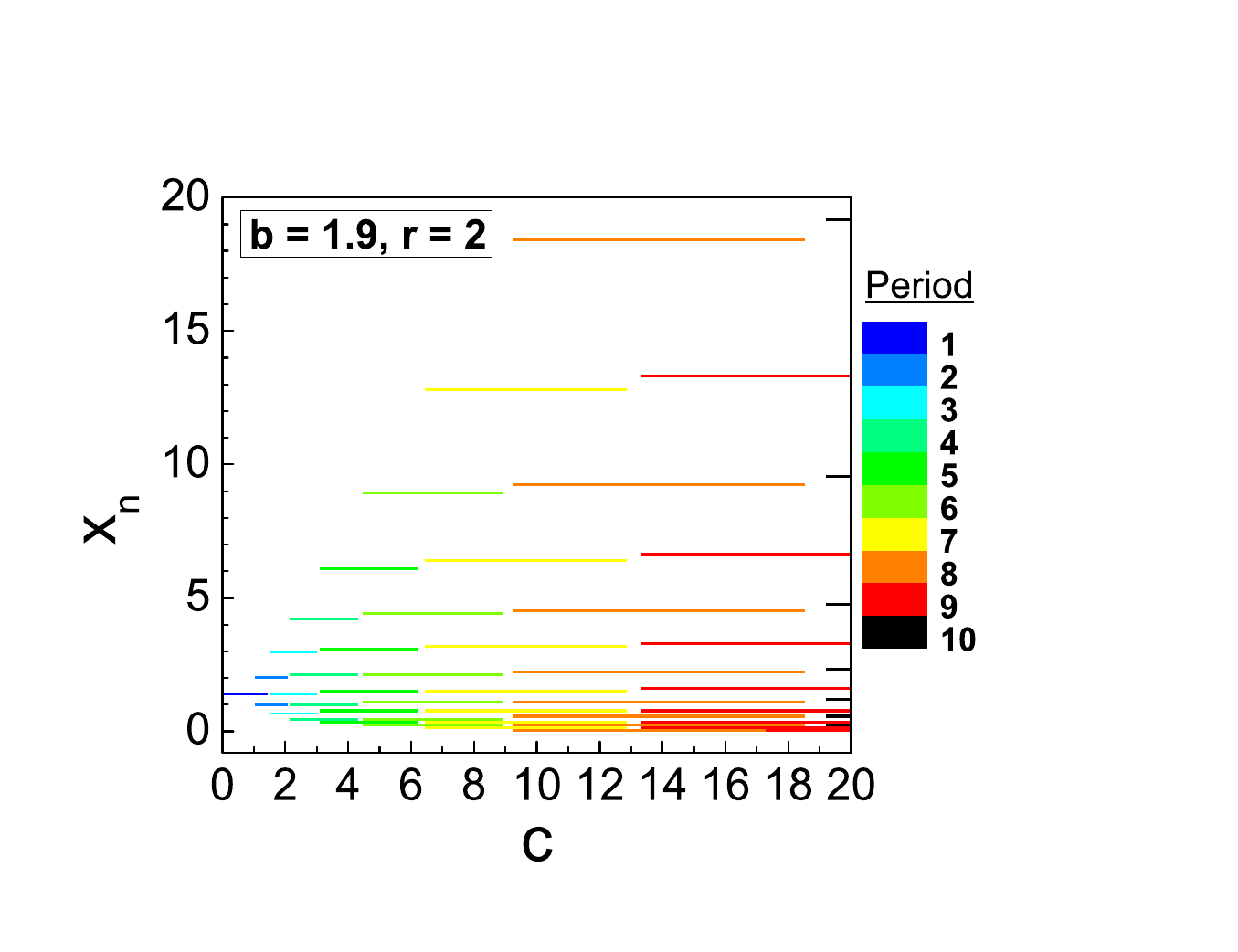}
\caption{\label{fig:seq}   (Color online) Regular orbits.
Bifurcation diagram  as a function of $c$, with $b=1.9$ and $r=2$.
The period of cycles is coded by the color scale (gray tone).}
\end{figure}

The diagram in Figure \ref{fig:seq} gives a perspective of this
situation for $b=1.9$ and $r=2$, as a function of $c$.

In the rest of this Section we focus our attention on the case $b=2$
which separates the regular and chaotic regimes. We present both
numerical and analytical results. The reader is referred to
\ref{app:bookkeeping} for details.

\subsection{Case $b=2$. \label{sec:x0ne0}}

For $b=2$ and fixed $r>1$ the orbit of an arbitrary initial
condition $x_0 \in [0,\infty)$, is simple enough to be rigorously
described. The behavior is varied and shows an interesting
dependence on the parameter $c$. In particular this permits to
appraise the effect of the discontinuity. As detailed below, the
general characteristic in this case is that all the initial
conditions are fixed, periodic or eventually periodic. There exist
an infinity of coexisting limit cycles. Their periods are allowed to
take only a  restricted set of values, which depend on the initial
condition and parameters value through very strict laws.  We defer
their detailed description to \ref{app:bookkeeping}. The reader can
find there precise information about the transients length and exact
account of the cycles. Here we report only the most salient features
in terms of the variable $x$ and threshold parameter $c$.

If $c \le 1$, then the closed interval $[c,r/c]$ is invariant under
the action of the map. Any interior point $x_0\ne \sqrt{r}$ belongs
to the $2-$cycle $\{x_0,r/x_{0}\}$. $x_0=\sqrt{r}$ is a fixed point.
The extremes $x_0=c$ and $x_0=r/c$ as well as all the exterior points are
eventually periodic, entering the invariant interval after a
transient.

If $c>1$, then the closed interval $[1/c,rc]$ is invariant under the
action of the map. Every exterior point is eventually periodic.
Interior points are periodic. For  fixed $c$, the period of the
cycles depends on $x_0$ but can take on only a restricted set of
values. For instance, with $c=2$ one finds a  $3-$cycle:
$\{1/\sqrt{r},\sqrt{r},r\sqrt{r}\}$; a $4-$cycle:
$\{1/r,1,r,r^{2}\}$; and the rest of the points in the interval
accommodates in an infinity of $6-$cycles. All these cycles are the
limit cycles for the eventually periodic points. Notice that in this
system the existence of period three does not imply chaos. This is
not in conflict with the celebrated Li and Yorke theorem because the
VGH map is defined by a discontinuous function.

The basins of attraction of such limit cycles are infinite
intermingled sets of zero measure made of equidistant points in the
$z$ variable. The vertical segments  located at $b=2$ in Figures
\ref{fig:V_r2_bif-b} and \ref{fig:V_C25_r2_xLy} comprise all the
coexisting limit cycles. In numerical simulations such a vertical
segment gets filled only if enough initial conditions are used in
the construction of the bifurcation diagram.

\subsection{An explanation of the observed bifurcation\label{sec:explica}}

\begin{figure}[H]
\includegraphics[scale=0.6]{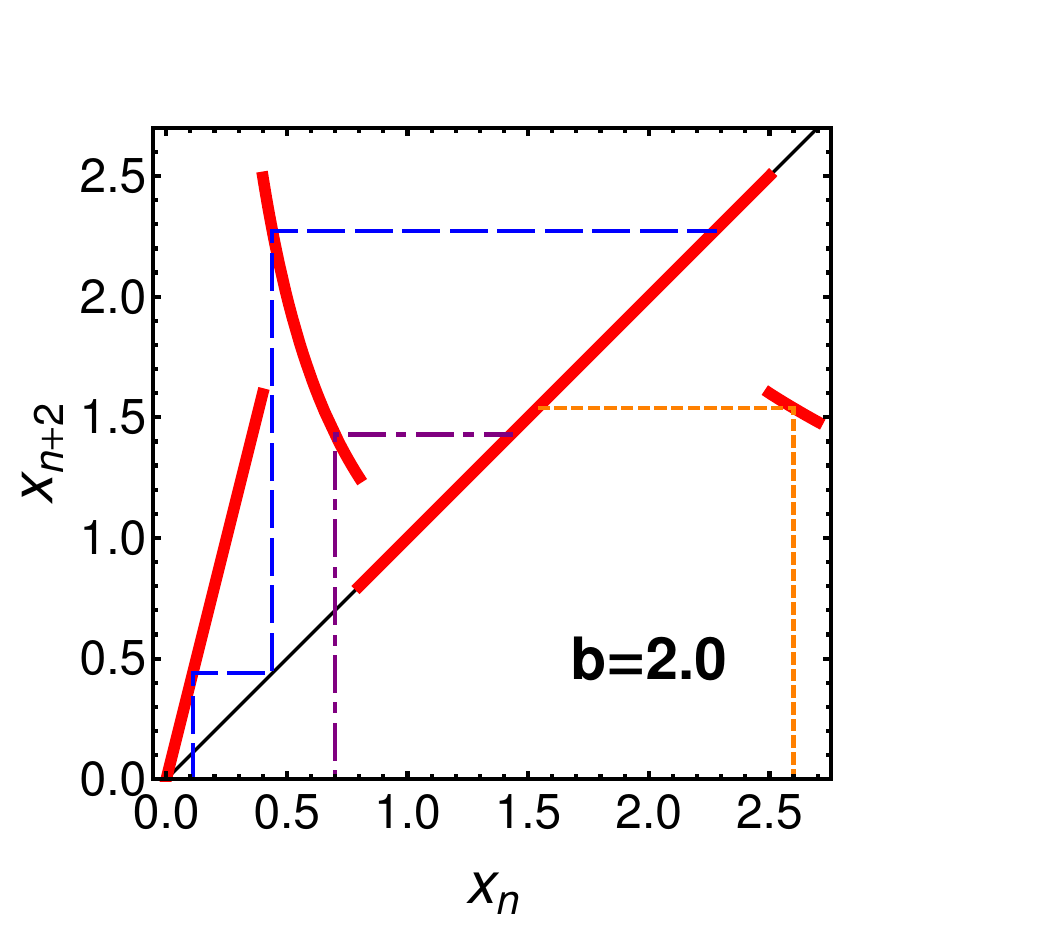}
\caption{\label{fig:cobweb2} (Color online) Cobweb plot for
$f^{[2]}$ (thick red line). Parameter values: $b=r=2,c=0.8$. The
three seeds are $x_0=0.11,0.7,2.6$, and are eventually stable fixed
points of $f^{[2]}$. }
\end{figure}

Next we develop an heuristic explanation for the origin of the
order-to-chaos bifurcation observed at $b=2$. As is well known,
period-2 points are fixed points of the second iterate $f^{[2]}$ of
the  map. In the cobweb plot of $f^{[2]}$ shown in Figure
\ref{fig:cobweb2} the relevant element is the piece of $f^{[2]}$
which is co-linear with the bisectrix, because it conveys the
existence of a continuous interval of fixed points. Leaving apart
the fixed point of $f$, these points pair themselves to form the
alluded period-2 orbits. By the cobweb construction the rest of the
points eventually land in the piece on the bisectrix.

Whenever an infinite set of trajectories of period $n>2$  is
observed, with $b=2$, it is because this value of the parameter
forces $f^{[n]}$ to have one or more segments co-linear with the bisectrix
too. The question is then: Which are the conditions for the
map $f$ to allow  iterates $f^{[n]}$ to have pieces
co-linear with the bisectrix?

Our answer is that the phenomenon may take place whenever the piecewise
continuous map
$f$ has a piece defined by a power-law function. To buttress this
answer let us write the general expression of a term of $f^{[n]}$
as
\begin{equation}\label{eq:genterm}
r^{p(b)} x^{(1-b)^{m}}
\end{equation}
where $p(b)$ stands for a polynomial of $b$ and $m\leq n \in
\mathbb{N}$. Since $b>1$, the conditions for (\ref{eq:genterm}) to be
co-linear with the bisectrix are: even $m$, $b=2$ and $p(2)=0$.

To further illustrate this, let us focus on one part of the second
iterate of (\ref{eq:Varley1}), namely
\begin{equation} \label{eq:iter2}
f^{[2]}(x)=\left\{
\begin{tabular}{ll}
      $\cdots ,$ & $\cdots $ \\
      $r^{2-b}x^{(1-b)^2},$ & $c<x<r/c$, $c<\sqrt{r}$
     \end{tabular}
     \right.
\end{equation}
where the dots stand for the three remaining terms and intervals in
the definition of the second iterate. The explicit term of $f^{[2]}$
written down in (\ref{eq:iter2}) is co-linear with the bisectrix in
the cobweb plot if and only if $b=2$, as expected, provided $c<\sqrt{r}$. If
$c\geq\sqrt{r}$ the extremes of the interval collapse to a single
point and that particular piece disappears in the second iterate.

\begin{figure}[H]
\includegraphics[scale=0.5]{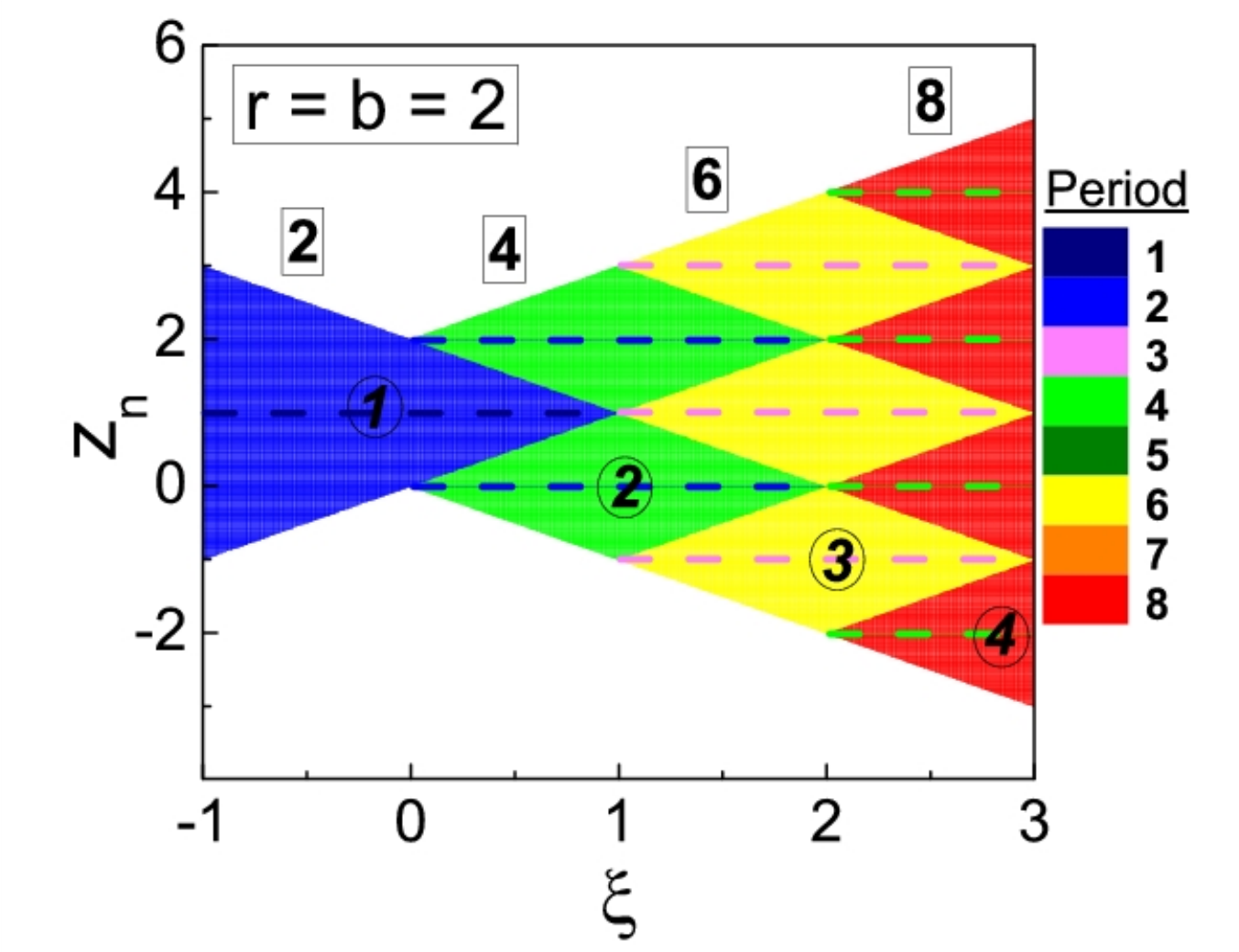}
\caption{\label{fig:peradding} (Color online) Bifurcation diagram for $r=b=2$.
The color (gray tone)
codes the period of the cycle which is also given by the framed numbers for trajectories
falling in rhombuses. The circled numbers stand for the periods of the cycles falling on
dashed lines. }
\end{figure}

In principle, several iterates can have co-linear terms
simultaneously for the same value of $c$. This gives rise to
coexisting cycles of different periods, a fact that is illustrated
in the bifurcation diagram  of Figure \ref{fig:peradding}. There we
have coded in color (gray tone) the periods of the trajectories,
with fixed values $b=r=2$. The representation is given in terms of
$\{ z,\xi\}$, which endows the plot with more symmetric structure.
The intermingling of the colored rhombuses accounts for the
coexistence of limit cycles with various periods. In addition,
horizontal dashed (color) segments stand for cycles whose periods
are indicated by circled numbers and have been taken from Table
\ref{tab:T}. Framed numbers indicate the corresponding period of
orbits in rhombuses. As Figure \ref{fig:peradding} shows, for the
parameter values considered, only two different iterates can have
co-linear terms for the same value of $\xi$. Moreover, as $\xi$
increases, more co-linear terms of successively higher iterates
emerge whereas the co-linear terms of the previous iterates collapse
and disappear (see \ref{app:bookkeeping}).

In other words, the presence of the $rx^{1-b}$ power-law piece in the definition
of the VGH model is at the origin of the bifurcation at $b=2$.
More generally, any one-dimensional map defined as a piecewise function,
with a power-law piece in its definition, is a candidate to exhibit a
bifurcation of the present type.

\subsection{Case $b=2$. When $x_0=c$: Harter's boundaries \label{sec:Harter}}
As is well known, for a continuous unimodal uniparametric map with
its maximum at $x=x^{*}$, the plot of the successive iterates of the
initial seed $x_0=x^{*}$ versus the map parameter generates the
so-called Harter boundaries \cite{Jensen,Eidson}. In the logistic
map \cite{May2, Collet, Feigen1, Feigen2}, for instance, they
correspond to the sharp cusps observed in the invariant density.
These boundaries are, to some extent, the skeletal system of the
bifurcation diagram where they are readily observed provided it is
built up using color or grey tones. Harter lines cross at unstable
equilibrium points and, in the case of the logistic map, the set of
crossing points follows itself a reversed bifurcation cascade.

The VGH map is not differentiable at $x=c$, although it is still
unimodal in the sense of being monotonically increasing for $x\le c$
and decreasing for $x>c$.  A similar study to the one described for
the logistic map leads to interesting results. To be precise, Harter
lines correspond to the color boundaries visible in Figures
\ref{fig:V_r2_bif-b} and \ref{fig:V_C25_r2_xLy}. They can be
directly established from the form (\ref{eq:Varley1}) of the map. It
is nevertheless simpler to use instead (\ref{eq:zlog}) and the
results gathered in \ref{app:lin}. For $b=2$, the seed $x_0=c$ ends
up in a cycle whose period $P(r,c)$ depends on $r$ and $c$. The
expression for $P$ and the cycle elements may be explicitly written
down.

For $c< 1$ the 2-cycle $\{rc, 1/c\}$ is reached just after one
iteration. For $c \ge 1$, $x=c$ is always a periodic point.
Obviously $c=1$ belongs to the 2-cycle $\{1,r\}$. When $c>1$ there
is a unique integer $M\in \mathbbm{N}$ such that
\begin{equation}
  c\in \left[
r^{M/2},r^{(M+1)/2}\right) .
\end{equation}
with
\begin{equation}
M=\left\lfloor 2\frac{\log c}{\log r} \right\rfloor =
\lfloor \xi \rfloor
\end{equation}
in terms of the integer part or floor function.  Then, for $c>1$
we get two cases
\begin{eqnarray}\label{eq:c-periodo2}
P(r,c)&=&M+2, \; \textrm{if}\;\; c=r^{M/2} \\
P(r,c)&=&2(M+2),\; \textrm{if}\;\; c\in \left(
r^{M/2},r^{(M+1)/2}\right) \label{eq:c-periodo3}
\end{eqnarray}
The $2(M+2)$-points of the cycle for the case (\ref{eq:c-periodo3})
read
\begin{equation}\label{eq:longperiod}
\left\{ \frac{c}{r^M},\frac{c}{r^{M-1}},\ldots,
\frac{c}{r},c,rc,\frac{1}{c},\frac{r}{c},\frac{r^2}{c},\frac{r^{M+1}}{c}
\right\}
\end{equation}
which in turn, for $c=r^{M/2}$ (\emph{i.e.}, case
(\ref{eq:c-periodo2})) contract to the $(M+2)$-points cycle
\begin{equation}
\left\{ r^{-M/2}, r^{-M/2+1},\ldots
,r^{M/2-1},r^{M/2},r^{M/2+1}\right\}
\end{equation}

It is interesting to observe that (\ref{eq:longperiod}) can be
obtained recurrently from $\{rc,1/c\}$ by a simple procedure. In
going from $M$ to $M+1$ two new elements are added to the cycle: the
one in the leftmost position is obtained by dividing by $r$ the
first element in the previous cycle. The other, which goes at the
rightmost position, is obtained by multiplying by $r$ the last
element of the previous cycle.

\section{Dependence on $c$: Discontinuity location \label{sec:xi}}
The dependence of the system with respect to the parameter $c$ with
$r$ fixed, or equivalently $\xi$, exhibits a large variety of
features. We describe the system in the chaotic regime. First near
the bifurcation point $b=2$, and next for higher $b$.

\subsection{Near $b=2$}\label{sec:near2}

\begin{figure}[H]
\includegraphics[scale=1.1]{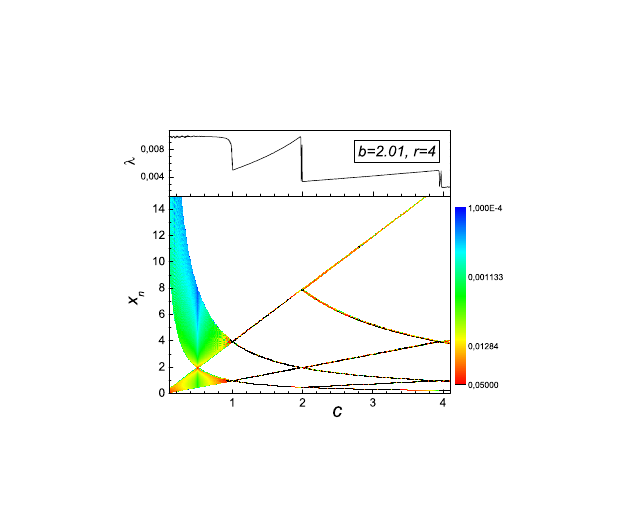}
\caption{\label{fig:V_b201_r4_bif_Ly}  (Color online) Bifurcation
diagram (bottom panel) and Lyapunov exponent (upper panel) obtained
using $b=2.01$ and $r=4$. }
\end{figure}

In Figure \ref{fig:V_b201_r4_bif_Ly} we plot the bifurcation diagram
as a function of $c$, computed with $b=2.01$ and $r=4$. For $c<1$
the trajectories $x_n$ wander in two relatively large chaotic bands
in contrast with the extremely narrow ones for $c>1$.

The upper panel in the figure gives the Lyapunov
exponent. It is everywhere positive, as it must be for a chaotic
regime. As a function of $c$, its value changes suddenly at every
band merging crisis observed in the bifurcation diagram.

\begin{figure}[H]
\includegraphics[scale=1.15]{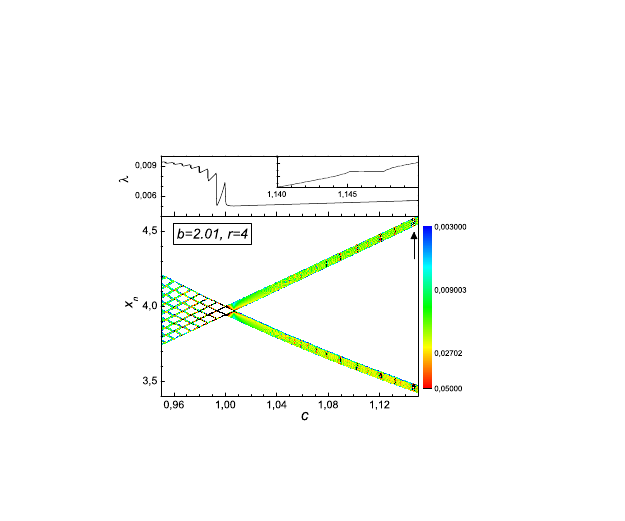}
\caption{\label{fig:V_b201_r4_bif_Ly_zoom}  (Color online)
Bifurcation diagram (bottom panel) and Lyapunov exponent (upper
panel) obtained using $b=2.01$ and $r=4$. This plot presents a zoom
of Figure \ref{fig:V_b201_r4_bif_Ly}. The inset in the upper panel
allows to appreciate the plateau in $\lambda$.}
\end{figure}

\begin{figure}[H]
\includegraphics[scale=1.1]{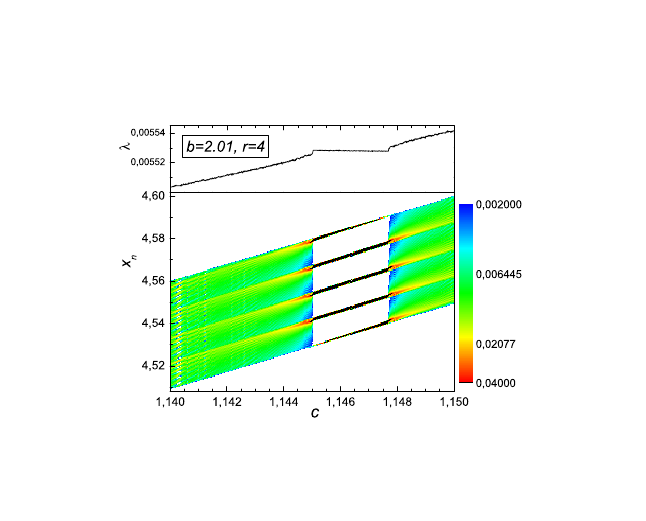}
\caption{\label{fig:V_b201_r4_bif_Ly_zoom-R}  (Color online)
Bifurcation diagram (bottom panel) and Lyapunov exponent (upper
panel) obtained using $b=2.01$ and $r=4$. This plot presents a zoom
of Figure \ref{fig:V_b201_r4_bif_Ly_zoom} in the region located by
the arrow. It exhibits interior crises.}
\end{figure}

\begin{figure}[H]
\includegraphics[scale=1.1]{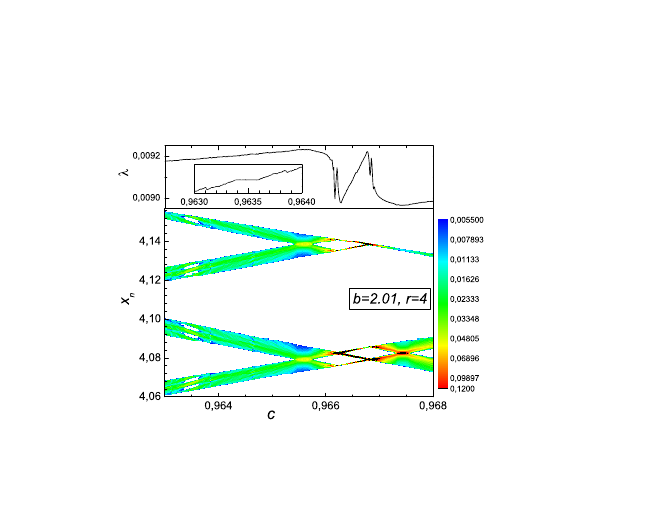}
\caption{\label{fig:V_b201_r4_bif_Ly_zoom-L}  (Color online)
Bifurcation diagram (bottom panel) and Lyapunov exponent (upper
panel) obtained using $b=2.01$ and $r=4$. This plot presents a zoom
of Figure \ref{fig:V_b201_r4_bif_Ly_zoom}. The inset in the upper
panel allows to appreciate the plateau in $\lambda$.}
\end{figure}

To study the diagram with further detail we present in
Figures \ref{fig:V_b201_r4_bif_Ly_zoom} --
\ref{fig:V_b201_r4_bif_Ly_zoom-L}  magnifications of small areas
in Figure \ref{fig:V_b201_r4_bif_Ly}.

The region with $c<1$, near $b=2$,  has further interesting
features. Thus, what in Figure \ref{fig:V_b201_r4_bif_Ly} appears as
two dense chaotic bands does have structure when minutely examined.
Figure \ref{fig:V_b201_r4_bif_Ly_zoom}  shows this feature for the
upper band. The grid structure in the bifurcation diagrams as a
function of $c$ (or $\xi$) fades as $c$ decreases.  The Lyapunov
exponent in the upper panels exhibits jumps associated with chaotic
bands merging crises.

\begin{figure}[H]
\includegraphics[scale=1.1]{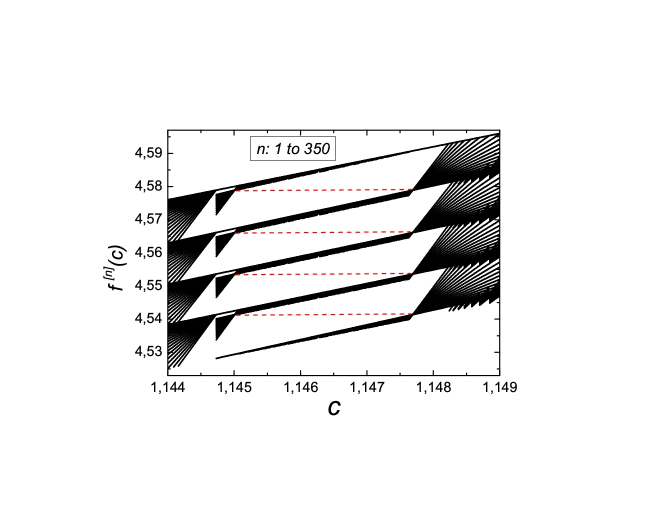}
\caption{\label{fig:Harter}  (Color online) Detail of the first 350
Harter curves obtained using $b=2.01$ and $r=4$. Dashed (red) lines
between crossings of harter curves correspond to unstable fixed
points.}
\end{figure}

For $c>1$, the bifurcation diagram in Figure
\ref{fig:V_b201_r4_bif_Ly_zoom} exhibits further details after
magnification. A zoom of the area indicated by the arrow is given in
Figure \ref{fig:V_b201_r4_bif_Ly_zoom-R} where a sudden variation of
the size of the attractor is apparent. This thin window is composed
by twenty two chaotic narrow bands (only five in the zoom). Further,
the crossing of Harter lines in Figure \ref{fig:Harter} punctuate
the start and end of the shrunk chaotic bands. A unstable orbit
exists inside the window, which is represented by the dashed lines
in the Figure \ref{fig:Harter}. Consequently, what we observe in
Figure \ref{fig:V_b201_r4_bif_Ly_zoom-R} as a shrinking and widening
of the attractor corresponds, actually, to a pair of interior crisis
\cite{Ott, Grebogi-Ott, Lai}. Moreover, further interior crises of
various smaller sizes appear inside the window itself in a,
possibly, self-similar way. For instance, in the leftmost part of
this bifurcation diagram ($1.14 < c < 1.1405$) one can hint a crisis
of the very same kind.

Figure \ref{fig:V_b201_r4_bif_Ly_zoom-L} corresponds to a zoom of
the left and uppermost part of the bifurcation diagram in Figure
\ref{fig:V_b201_r4_bif_Ly_zoom}. The Lyapunov exponent in the upper
panel shows oscillations at band merging crises and, once again, a
plateau (only visible in the inset) at the crisis located at
$0.963<c<0.964$. An explanation for the occurrence of $\lambda$
plateaus is deferred to Section \ref{sec:plateaus}.

\begin{figure}[H]
\includegraphics[scale=1.1]{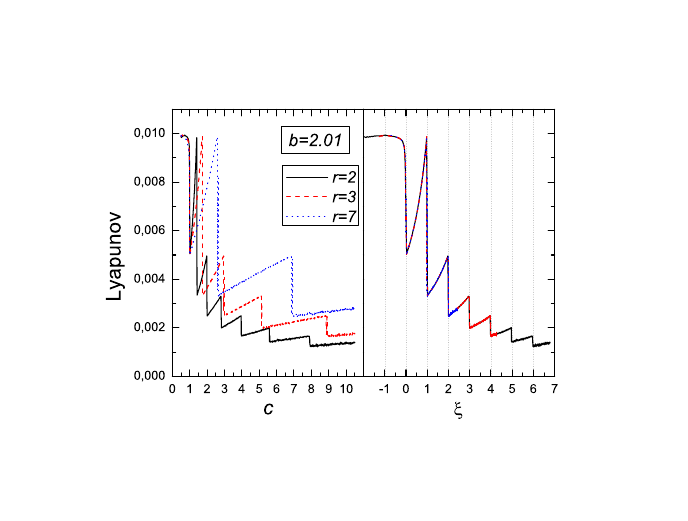}
\caption{\label{fig:V_Ly_collapse} (Color online) Left panel:
Lyapunov exponent as a function of $c$ for three values of $r$.
Right panel: Collapse of the same three curves as a function of
$\xi$}
\end{figure}

For $c>1$ and close to the critical point $b=2$,  the crises take
place at integer values of $\xi$. This feature is illustrated in
Figure \ref{fig:V_Ly_collapse}. The left panel shows $\lambda$ as a
function of $c$ for $b=2.01$ and three different values of $r$. The
right panel corresponds to the same data expressed in terms of
$\xi$: All three curves collapse onto a master curve. Furthermore,
the value of the Lyapunov exponent is invariant also in magnitude.
This property fades as far as we move toward higher values of $b$,
namely far from the critical point. All this buttresses the
existence of universality in the system near the transition
order-to-chaos.

Bifurcation diagrams of the VGH system always collapse when
expressed in terms of $z$ versus $\xi$. This is true even far from
$b=2$, in contrast with the Lyapunov exponent diagrams.

\subsection{Far from $b=2$}\label{sec:far2}

\begin{figure}[H]
\includegraphics[scale=1.0]{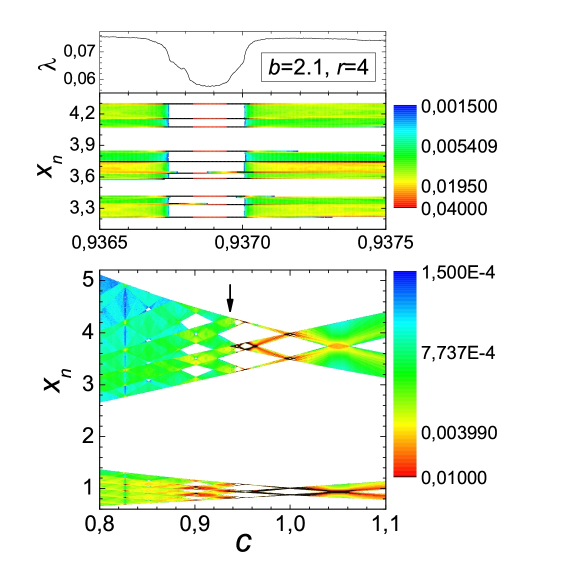}
\caption{\label{fig:V_c_b21_r4}  (Color online) Bottom panel:
Bifurcation diagram as a function of $c$, with $b=2.1$ and $r=4$.
The arrow locates the narrow window zoomed in the upper panel
where $\lambda$ is also shown. }
\end{figure}

The lower panel of Figure \ref{fig:V_c_b21_r4} gives a view of the
bifurcation diagram near $c=1$, with $r=4$ and $b=2.1$, i.e. leaving
the critical point. The upper panel shows an enlargement of the
region located by the arrow below. This is again an instance of
interior crisis. Concomitantly, the uppermost panel shows the
variation of the Lyapunov exponent. At variance with Figures
\ref{fig:V_b201_r4_bif_Ly_zoom}--\ref{fig:V_b201_r4_bif_Ly_zoom-L},
here $\lambda$  exhibits neither jumps nor plateaus across the
window where the attractor shrinks. The reason is explained in the
next subsection.

\begin{figure}[H]
\includegraphics[scale=1.1]{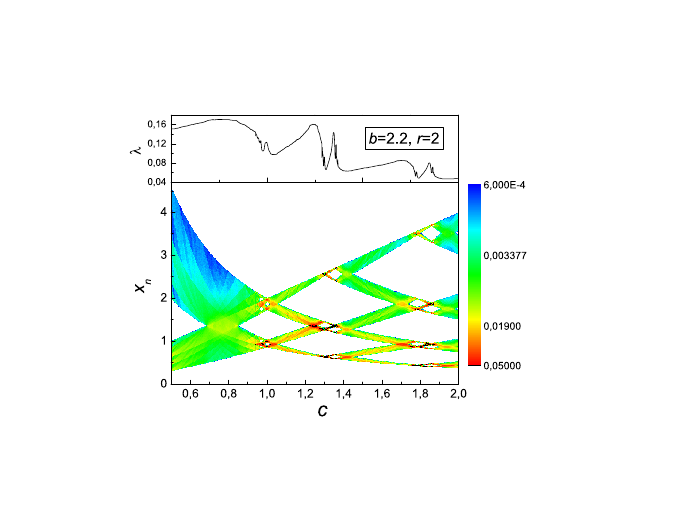}
\caption{\label{fig:V_c_b22_r2} (Color online) Bifurcation diagram
and Lyapunov exponent as a function of $c$. This particular plot was
obtained using $b=2.2$ and $r=2$. The system is not close to the
critical point $b=2$.}
\end{figure}

In the bifurcation diagram of Figure \ref{fig:V_c_b22_r2}, with
$b=2.2$ and $r=2$, the band merging phenomenon is more involved than
in Figure \ref{fig:V_b201_r4_bif_Ly}. The corresponding Lyapunov
exponent, in the upper panel, exhibits large variations at crises
too. However, these do not take place any longer at integer values
of $\xi$.

\begin{figure}[H]
\includegraphics[scale=1.1]{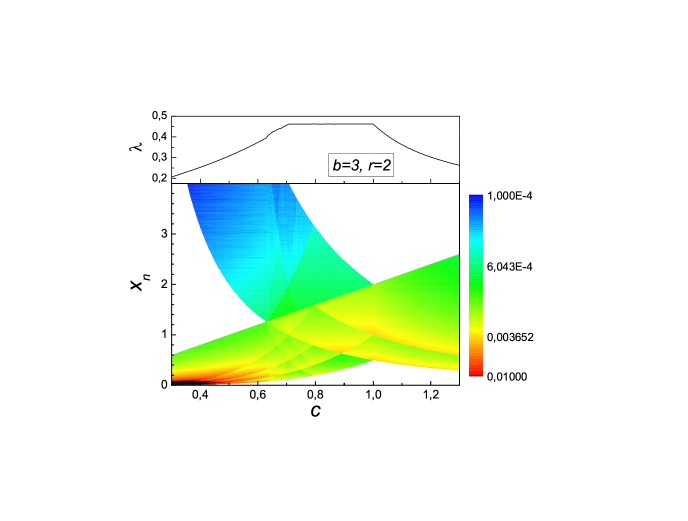}
\caption{\label{fig:V_b3_bif-L}  (Color online) Bifurcation diagram
and Lyapunov exponent as a function of $c$, with $b=3$ and $r=2$. }
\end{figure}

Eventually, in Figure \ref{fig:V_b3_bif-L}, where $b=3$, we observe
that the Lyapunov exponent presents a large plateau whereas in the
corresponding bifurcation diagram no interior crisis occurs. We have
not found a justification for this case.

\subsection{Interior crises and $\lambda$ plateaus}\label{sec:plateaus}

An plausible explanation for the origin of $\lambda$ plateaus
occurring between interior crises pairs reads as follows. According
to our numerical simulations, the narrow windows between interior
crises of the VGH map resemble very much the so-called cycles of
chaotic intervals, where the trajectories jump in a cyclic way from
one chaotic band to another. Such a phenomenon occurs, for instance,
in the logistic map \cite{Schlogl} and is also termed in the
literature as cyclic chaotic attractor or cyclic chaotic bands
\cite{Maistrenko}.

Trajectories look quite regular and, if the line $x_n=c$ in the
bifurcation diagram does not cross any of the thin bands inside the
window, then the proportion of exterior points ($x_n>c$) remains
almost constant. As a consequence of it and taking into account the
interpretation of $\lambda$ for this map at the end of Section
\ref{sec:Lyap}, the Lyapunov exponent gets the plateau shape.
Otherwise, $\lambda$ varies across the window, which is the case in
Figure \ref{fig:V_c_b21_r4}.

\subsection{Crisis-induced intermittency \label{sec:traj}}

Next we gather some results, obtained from numerical simulations,
concerning the behavior of trajectories around interior crises.

We commence by pointing out that interior crises in the VGH map
appear in pairs. As the parameter $c$ increases through a star value
$c^*_1$ the chaotic attractor suddenly shrinks into thin chaotic
bands. This is the first interior crisis. Then, it exists a second
star value $c_2^*>c_1^*$ where the set of thin bands suddenly widen
and the attractor at $c>c^*_2$ recovers its old size. This is the
second interior crisis of the pair. By contrast, in the logistic
map, a tangent bifurcation precedes always an interior crises
\cite{Ott}.

\begin{figure}[H]
\includegraphics[scale=1.0]{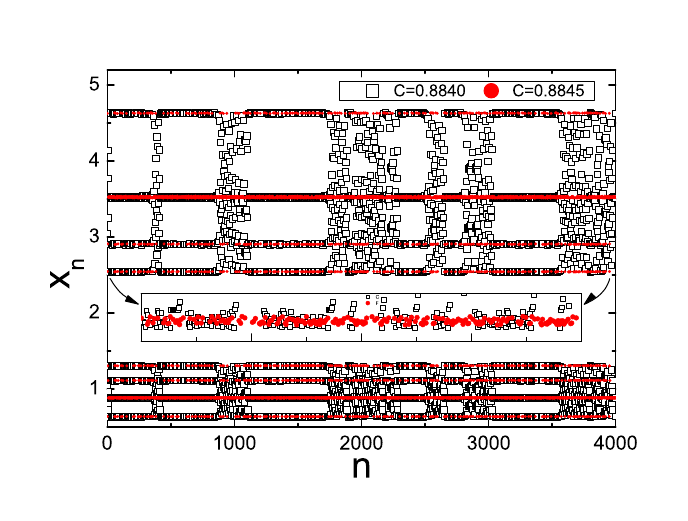}
\caption{\label{fig:V_intermit} (Color online)  Crisis-induced
intermittency, with $r=4$ and $b=2.2$. Two trajectories from both
sides of the critical value $c^*=0.88411175\ldots $ are compared.
Solid dots (red) correspond to a trajectory with $c=0.8845$, where
the attractor splits in narrow bands. Empty squares (black)
correspond to $c=0.8840$, namely in the wider region of the
attractor. The inset is a zoom of the points close to $x=2.5$ and
allows to illustrate the non-regular character of these
trajectories. }
\end{figure}

For values of $c$ slightly different than a star value, in the
region where the attractor widens ($c\lesssim c_1^*,c\gtrsim
c_2^*$), the orbits spend long stretches in the region where the
attractor is confined between the two crises. Occasionally, the
trajectories burst and visit the whole attractor. This behavior,
termed crisis-induced intermittency \cite{Ott, Romeiras}, is
illustrated in Figure \ref{fig:V_intermit}. There, two orbits are
plotted for slightly different values $c_1=0.8840$ and $c_2=0.8845$,
located at both sides of the star value $c^*=0.88411175\ldots$. The
empty square symbols stand for a trajectory in a regime where the
attractor is still large. Intermittency is clearly observed. The
solid dots represent a trajectory in a regime with shrunk attractor.
It seems, at first glance, a period-8 orbit but the inset, which is
a zoom of just one single narrow band, allows us to illustrate its
non-periodic character.

\begin{figure}[H]
\includegraphics[scale=0.3]{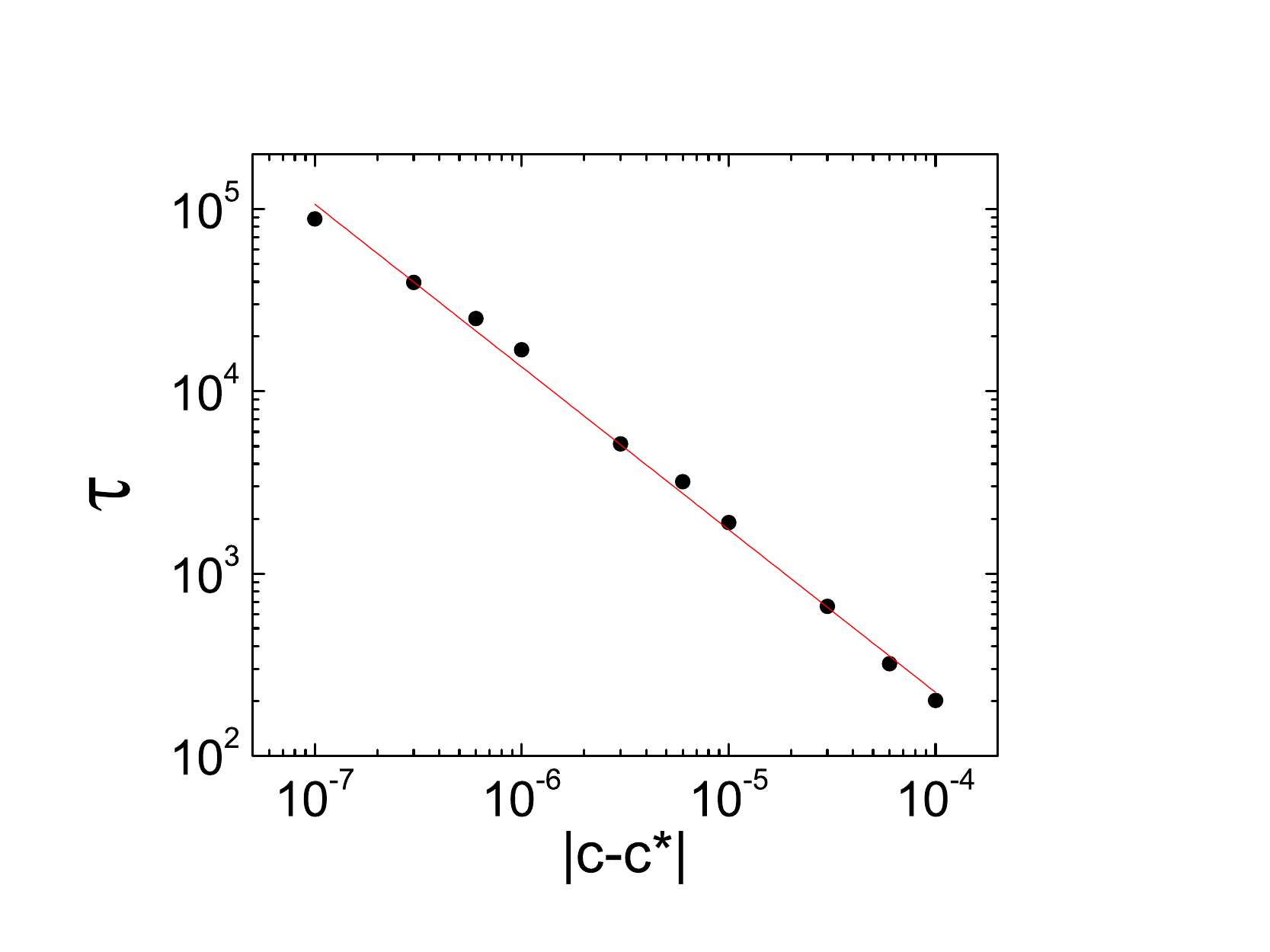}
\caption{\label{fig:tauvsdist}  (Color online) Characteristic length
$\tau$ versus $|c-c^*|$ in log-log scales. Each dot was obtained
from 100 initial conditions and series of $10^6$ iterates, with
$c^*=0.88411175$, $b=2.2$ and $r=4$. The slope of the linear fit is
$\gamma= -0.89\pm 0.02$.}
\end{figure}

For fixed $c$, the statistics of the length of stretches where the
orbit stays confined in the region of the shrunk attractor is well
described by an exponential distribution, provided $|c-c^*|$ is
small, which yields a characteristic length $\tau (c)$. It has been
shown \cite{Romeiras} that for a large class of dynamical systems
which exhibit crises, the dependence reads
\begin{equation} \label{eq:powerlaw}
\tau \sim |c-c^*|^{-\gamma}.
\end{equation}
Indeed, this is the case for the VGH map. We have built up the
statistics of $\tau$ as a function of $|c-c^*|$ for the case in
Figure \ref{fig:V_intermit}. The results are given in Figure
\ref{fig:tauvsdist}. The linear fit in log-log scales yields a
determination of the critical exponent: $\gamma=0.89\pm 0.02$.

\subsection{A numerical flaw \label{sec:b-integer}}

Version (\ref{eq:zlog}) of the map presents a worth mentioning
numerical nuisance. Namely, for some integer values of $b>2$,
computer generated trajectories collapse to an unstable periodic
orbit after some iterations. A similar nature numerical artifact has
been studied in \cite{Diamond, Yuan}.

The dynamics for real $z$ may be viewed as follows.
For an initial point $z_0<\xi$, every iteration conveys a shift to
the right by two units until the condition $z_n>\xi$ is fulfilled.
The second line in (\ref{eq:zlog}) may be read as: $\{(1-b)\lfloor
z_n \rfloor +2\} +(1-b)(z_n \;\mathrm{mod} \;1)$. Thus, for integer
$b>2$ the quantity in brackets is an integer and hence, its
successive iteration conveys just a shift, as above. The key point
is that, for finite machine precision, the last term dramatically
looses precision in each iteration for some particular integer
values of $b>2$.

This effect may be understood on the basis of a generalization of
the Bernoulli or binary shift map: $z_{n+1}=2z_n\; (\mathrm{mod} \;
1)$. To this end, let us use the binary representation of the number
$z_n \mathrm{mod} \;1$. Multiplication in base $2$ is very simple in
some cases. For instance, when $|1-b|=2^{k}$ we get the binary
representation of the product $2^k (z_n\; \mathrm{mod} \;1)$ simply
by shifting $k-1$ places to the left every bit and (due to the
finite precision of the computer) adding simultaneously $k-1$ zeros
to the right of the number. Henceforth this number gets shifted by
two units in every iteration till it reaches $z_n>\xi$ when the
$\mathrm{mod}$ operator  acts again. This way, the piece in bracket
is preserved as an integer under the action of the floor function,
whereas the term $2^k (z_n \; \mathrm{mod} \;1)$ looses significant
digits before going through the loop again. Eventually it stops
when, after a number of iterations, $2^k (z_n\; \mathrm{mod}
\;1)=0$. At this point the iteration reduces to a periodic orbit
with elements located at integer numbers. The full output is then a
set of limit cycles whose elements are always integers. This
misleading result is just consequence of the computer finite
precision.

\section{Discussion and conclusions \label{sec:fin}}

We have revisited a one-dimensional population model, proposed as an
instance of density dependent dynamics. The fitness parameter $b$
controls the onset of chaos. Values $b>2$ convey chaos. Otherwise
the system is regular. On the critical point $b=2$ the attractor is
made of an infinity of limit cycles that share a finite number of
different periods. Hence, the transition order-to-chaos takes place
through three steps: \emph{i}) Finite number of limit cycles at
$b<2$; \emph{ii}) Infinite number of limit cycles with finite number
of different periods at $b=2$; and \emph{iii}) Chaos at $b>2$. No
period doubling occurs in this route to chaos. At $b=2$, we have
given an exact description of the system. We have checked that such
regular attractors have been sometimes plotted in bifurcation
diagrams in the literature. However we have not found the concurrent
descriptions of the attractor. Moreover, bifurcation diagrams may be
found in the literature where this attractor lacks, most likely
because too few initial seeds were used in the computation. An
explanation for the phenomenon to occur has been provided on the
basis of the power-law tail of the map.

The behavior of continuous piecewise linear maps has been studied in
terms of the so-called border collision bifurcations. Thus, a
thorough classification of the different kinds of bifurcations, in
the parameter space, is given in \cite{Banerjee1}. The study
explicitly excludes the borderline systems where a slope equals $\pm
1$. Interestingly, the VGH map in its version (\ref{eq:zlog}) with
$\xi=0$ $(c=1)$, falls just into this category. A study of
discontinuous maps \cite{Avrutin3}, namely with $\xi\ne 0$ (or $c\ne
1$), is flawed by the same restriction. Hence, the VGH map
corresponds to a family of functions that are, by definition, out of
those analyses. Moreover, in view of the results given in Section
\ref{sec:explica}, this type of bifurcation may endow that
classification with further structure.

An interesting phenomenon occurs close to the critical value, i.e.
for $b\simeq 2$. There, the Lyapunov exponent of the VGH model
exhibits renormalization features when expressed as a function of
the variable $\xi$. In turn, the bifurcation diagrams as a function
of $c$ collapse for all $r$ and $b$ when expressed as $\{z_n,\xi\}$.

The abrupt variation of the Lyapunov exponent near band merging
crises has already been reported \cite{Mehra}. Moreover, we have
observed plateaus in the Lyapunov exponent, when represented as a
function of $c$. It is worth mentioning that interior crises in the
VGH map appear always in pairs.

We think that the the presence of a power-law piece in maps like the
VGH model studied here originates worth knowing properties. We do
hope that the variety of results gathered in this work may stimulate
further work.

{\emph{Acknowledgments.}} This work has been partially supported by
contracts MCyT/FEDER, Spain (Grant No. FIS2004-0912) and Generalitat
Valenciana, Spain (Grant No. ACOMP07/03). VBS thanks Generalitat
Valenciana for financial support.

\appendix
\section{Exact piecewise linearization \label{app:lin}}

Although by no means necessary, version (\ref{eq:zlog}) of the VGH
dynamical system is extremely useful for analytic purposes and has a
simple mathematical interpretation: it corresponds to two coupled
affine dynamical systems. Precisely the new parameter $\xi$
introduced acts as coupling constant.

For the general affine discrete dynamical system
\begin{equation} \label{eq:affine1}
  x_{n+1}=Ax_{n}+B
\end{equation}
the general solution with initial condition $x_{0}$ is
\begin{equation} \label{eq:affine}
x_{n}=\left\{
\begin{tabular}{ll}
      $x_0 + Bn,$ & \textrm{if}  $A=1$ \\
      $\left( x_0+\frac {B}{A-1}\right) A^{n}-\frac{B}{A-1},$ & \textrm{if}  $A\ne 1$
     \end{tabular}
     \right.
 \nonumber
\end{equation}
By itself system (\ref{eq:affine1}) has a rather dull dynamics. The
evolution tends to the fixed point ${B}/{(1-A)}$ if $|A|<1$, whereas
escapes to infinity if $|A|>1$ or $A=1$. For $A=-1$ any initial
condition $x_0$ except the fixed point ${B}/{2}$ belongs to the
2-cycle $\{x_0,-x_0+B\}$.

For the system (\ref{eq:zlog}), the coupling parameter $\xi$ divides the
phase space for our variable z in two regions: $z\le \xi$ (Region I)
and $z> \xi$ (Region II). In I (\ref{eq:affine1}) holds with $A=1$,
whereas $A=1-b<0$ in region II. All over the phase space $B=2$. The
previous equation (\ref{eq:affine}) is instrumental in discussing
the transitions to and fro between both regions. These transitions
explain the more complicated dynamics of the coupled system
(\ref{eq:zlog}).

Transitions $I\rightarrow II$ from a point $z_0 \le \xi$ will
necessarily take place after $N$ iterations with
\begin{equation}\label{n12}
N=\left\lfloor \frac{\xi-z_0}{2} \right\rfloor+1
\end{equation}
independently of the value of parameter $b$. Furthermore the orbit
of $z_0 \le \xi$ will land in II always with $z_N \in (\xi,\xi
+2]$.

In contrast, if $z_0> \xi$  transitions $II\rightarrow I$ are not
always possible and, when they actually occur, follow a more
complicated pattern depending on $b, \xi$ and $z_0$. Concerning
parameter $b$ it is convenient to distinguish two possibilities:
\begin{equation}\label{eq:AB}
\begin{tabular}{ll}
  $A:$ & $1 < b \le 2$ \\
  $B:$ & $b>2$ .\\
\end{tabular}
\end{equation}

As $\xi$ and $z_0$ are concerned we set apart three cases:
\begin{equation}
\begin{tabular}{lll}
  a: & $\xi < z_0\le 0$ \\
  b: & $\xi\le 0<z_0$ \\
  c: & $0<\xi<z_0$ .\\
\end{tabular}
\end{equation}

We have the following variants for case $A$ in (\ref{eq:AB}):
\begin{enumerate}
  \item [Aa:] transitions $II\rightarrow I$ are forbidden.
  \item [Ab:] transitions allowed only if $z_0>{(\xi-2)}/{(1-b)}$ and
  then   $N=1$
  \item [Ac:] transitions allowed for any $z_0$ if $\xi>2/b$ or if
  $\xi<2/b<z_0$ and $z_0>(\xi-2)/(1-b)$. In any case $N=1$.
  \end{enumerate}
  To discuss case $B$ in (\ref{eq:AB}) we introduce
  \begin{equation}
  \rho=\frac{\ln|(b\xi-2)/(bz_0-2)|}{\ln(b-1)}.
  \end{equation}
  We have the following variants:
  \begin{enumerate}
  \item [Ba:] transitions are always allowed and $N=2
  \lceil\rho/2\rceil$, in terms of the ceiling function.
  \item [Bb:] If $z_0<2/b$ then
  $N=2\lceil{\rho}/{2}\rceil$. For
  $z_0>2/b$, if $z_0<4/b-\xi$ then $N=\lfloor\rho\rfloor +1$ or $N=\lfloor\rho\rfloor +2$
  according to $ \lfloor\rho\rfloor$ being even or odd. If
  $z_0>2/b$ and $z_0>4/b-\xi$ then $N=1.$
  \item [Bc:] the trajectory always crosses to region I with varying
  N according to the following scheme:
\begin{enumerate}
  \item For $\xi<z_0<2/b$, $N=2
  \lceil{\rho}/{2}\rceil$

  \item For $\xi<(2/b)<z_0<(4/b-\xi)$, then $N=\lfloor\rho\rfloor+1$ if
  $\lfloor\rho\rfloor$ is even, or $N=\lfloor\rho\rfloor+2$ if $\lfloor\rho\rfloor$ is odd.
  \item For $\xi<2/b<4/b-\xi<z_0$, $N=1$
  \item For $2/b<\xi$, $N=1$, independently of $z_0>\xi$
\end{enumerate}

\end{enumerate}

\section{Detailed study of the system $b=2$}\label{app:bookkeeping}

Here we obtain the basins of attraction, periods, cycles and
transients for the VGH map with $b=2$ and arbitrary $\xi$ and $x_0$.

We analyze separately the cases $\xi <0$ ($ c< 1$), $\xi=0$ ($c=1$)
and $\xi >0$ ($c>1$) with fixed $b=2$. To facilitate the notation,
in this Appendix we will use the $z-$version of the map, namely
\begin{equation}
z_{n+1}=\left\{
\begin{tabular}{ll}
      $z_n + 2,$ & $z_n\le \xi$ \\
      $-z_n +2,$ & $z_n>\xi$
     \end{tabular}
     \right.
 \nonumber
\end{equation}

\subsection{$\xi <0$ (or $c<1$)}
  The first general feature of the
  system is that the interval $[\xi,2-\xi]$ is invariant under the
  action of the map. This is
  readily appreciated on the bifurcation diagrams in figures \ref{fig:V_r2_bif-b}
  and \ref{fig:V_C25_r2_xLy}.
  Every initial $z_0$ in the invariant interval belongs to a cycle of period two,
  with the exception $z_0=1$ which is a
  fixed point. The point $x_0=0$ (or equivalently $z_0=-\infty$) is
  fixed $\forall c$ (respectively $\forall \xi$).
  The remaining points $z_0$ are eventually periodic and, after a
  transient, they enter the interval $[\xi,2-\xi]$. The $z-$phase
  space get partitioned as
\begin{equation}\label{parti1z}
    (-\infty,\infty)=(-\infty,\xi]\bigcup(\xi,2-\xi)\bigcup[2-\xi,\infty),
\end{equation}
which corresponds to the original $x-$phase space partition
\begin{equation}\label{parti1}
    (0,\infty)=(0,c]\bigcup(c,\frac{r}{c})\bigcup[\frac{r}{c},\infty).
\end{equation}
  The dynamics of the map encompasses three cases, defined by the value of $\xi$:
\begin{enumerate}
  \item $z_0\le \xi$ (or $x_0\le c$). Define $N\equiv\lfloor {(\xi-z_0)/}{2} \rfloor$.
  Transient: $N+1$.
  Limit cycle: $\{ 2(N+1)+z_0,-2N-z_0\}$.
  \item $\xi<z_0 <2-\xi$ (or $c<x_0<r/c$). No transient. Cycle: $\{
  z_0,2-z_0\}$. $z_0=1$ fixed point.
  \item $2-\xi \ge z_0$ (or $r/c \ge x_0$). Define $N\equiv\lfloor {(\xi+z_0)}{2}\rfloor
  $. Transient: $N+1$.
  Limit cycle: $\{ 2(N+1)-z_0,-2N+z_0\}$.
\end{enumerate}

\subsection{$\xi =0$ (or $c=1$)}
  The invariant interval under the action of the map
  is $[0,2]$,
  and their points belong to cycles of period two, with the
  exception $z_0=1$ which is a fixed point. Points $z_0 \notin
  [0,2]$ are eventually periodic.
  The partition of the $z-$phase space is
\begin{equation}\label{parti0z}
    (-\infty,\infty)=(-\infty,0)\bigcup[0,2]\bigcup(2,\infty),
\end{equation}
which corresponds to the original $x-$phase space partition
\begin{equation}\label{parti11}
    (0,\infty)=(0,1)\bigcup[1,r]\bigcup(r,\infty).
\end{equation}
  We distinguish four cases:
  \begin{enumerate}
    \item $z_0< 0$ (or $x_0<c=1$). Define $N\equiv -\lfloor z_0/{2} \rfloor$.
    Transient: $N$. Cycle: $\{ 2N+z_0,-2(N+1)-z_0\}$.
    \item $z_0=0$ and $z_0=2$ constitute the cycle: $\{ 0,2 \}$.
    \item $0<z_0 < 2$ with $z_0\ne 1$.
    Cycle: $\{ z_0,2-z_0\}$.
    \item $z_0 >2$ (or $x_0>r$). Define $N\equiv \lfloor z_0/{2} \rfloor$.
    Transient: $N+1$. Limit cycle: $\{ 2(N+1)-z_0,z_0-2N\}$.
  \end{enumerate}

\subsection{$\xi >0$ (or $c>1$)}
  This is the most involved situation. Here
  the invariant interval under the action of the map
  is $[-\xi,2+\xi]$. The remaining
  points are eventually periodic. The convenient partition of the
  $z-$phase space is
  \begin{equation}\label{parti2z}
    (-\infty,\infty)=(-\infty,-\xi)\bigcup[-\xi,2+\xi]\bigcup(2+\xi,\infty).
  \end{equation}
  In the original $x-$phase space we
  have the partition
  \begin{equation}\label{parti2}
    (0,\infty)=(0,\frac{1}{c})\bigcup[\frac{1}{c},rc]\bigcup(rc,\infty).
  \end{equation}
  Next we describe the features of the various intervals:
\begin{enumerate}
  \item $z_0 < -\xi$ (or $x_0< 1/c$). Transient: $\lfloor -(\xi+z_0)/{2}
  \rfloor +1$.
  \item $z_0 > 2+\xi$ (or $x_0 > rc$). Transient: $\lfloor (z_0-\xi)/{2}
  \rfloor +1$.
  \item $-\xi \le z_0 \le 2+\xi$ (or $1/c \le x_0 \le rc$). This case
  embraces five possibilities according to the value of $\xi$:
  \begin{enumerate}
    \item $0<\xi <1$. The convenient partition of the $z-$phase
    space is
  \begin{eqnarray}\label{parti3}
    &[-\xi,2+\xi]=& \\ &[-\xi,0)\bigcup[0,\xi]\bigcup(\xi,2-\xi)& \nonumber\\
    &\bigcup[2-\xi,2)\bigcup[2,2+\xi] & \nonumber\\
    &\equiv \mathcal{A \bigcup B \bigcup I \bigcup C \bigcup D}.&
    \nonumber
  \end{eqnarray}
  The subinterval $\mathcal{I}=(\xi,2-\xi)$ is, in turn, invariant under the
  action of the map. $z_0=1$ is a fixed point. Else, orbits are
  the period-2 cycles: $\{ z_0,2-z_0\}$. The other four subintervals
  together form also an invariant subset. Their points hop
  in period-4 trajectories following the symbolic cyclic
  sequence: $\mathcal{A \to C \to B \to D}$.
    \item $\xi=1$. The interval under scrutiny is now $(-1,3)$. There exist
    the cycles: $\{-1,1,3\}$ and $\{0,2\}$. Points $z_0\ne -1,0,1,2,3$ are in
    cycles of period four.
    \item $1<\xi <2$. The convenient partition here reads:
  \begin{eqnarray}\label{parti4}
    & [-\xi,2+\xi]=& \\ & [-\xi,\xi-2]\bigcup(\xi-2,0)\bigcup \{0\}& \nonumber\\
    &\bigcup(0,2-\xi)\bigcup[2-\xi,\xi]\bigcup(\xi,2) & \nonumber \\
    &\bigcup \{2\}\bigcup(2,4-\xi)\bigcup[4-\xi,2+\xi] & \nonumber\\
    &\equiv \mathcal{A \bigcup B \bigcup C \bigcup D \bigcup E \bigcup F \bigcup
    G}. & \nonumber
  \end{eqnarray}
  There is no invariant subinterval under the map action.
  Four different types of periodic orbits exist. One period-2 trajectory which is the cycle
  $\{0,2\}$.  One period-3 trajectory which is the cycle $\{-1,1,3\}$. Period-4
  orbits are, symbolically: $\mathcal{B\to E \to C \to F}$, cyclic.
  Similarly, period-6 orbits follow the pattern:\\ $\mathcal{A\to D \to G \to A\to D \to G}$, cyclic.
  Notice that every one of the three subintervals is visited twice
  in one cycle.
    \item $\xi=2$. One period-3 cycle: $\{-1,1,3\}$. One period-4
    cycle: $\{-2,0,2,4\}$. The remaining orbits are period-6.
    \item $\xi >2$. This is by far the most involved situation.
    We define the following quantities: $N\equiv \lfloor
    \xi/2\rfloor$ and $\alpha\equiv \xi-2N$. The convenient
    partition now depends on $N$ and $\alpha$:
    \begin{eqnarray}
    &[-\xi,2+\xi]=&\\ &J_N^-\bigcup H_{N-1}^-\bigcup J_{N-1}^- \bigcup
    \ldots \bigcup H_0^- &\nonumber\\
    &\bigcup J_0^- \bigcup J_0^+ \bigcup H_0^+ \bigcup J_1^+ \bigcup H_1^+ & \nonumber \\
    &\bigcup \ldots\bigcup J_N^+ \bigcup H_N^+ \bigcup J_{N+1}^+ & \nonumber \\
    &\bigcup  \mathcal{A} \bigcup \mathcal{B} \bigcup \mathcal{C} \bigcup
    \mathcal{D}. & \nonumber
    \end{eqnarray}
    The length of the subintervals $J$'s and $H$'s is $\alpha$ and $2-\alpha$
    respectively. The super-index tells whether the interval is located either to the left
    ($-$) or to the right ($+$) of $z=0$. The sub-index codes the
    borders location of the interval according to
    \begin{eqnarray}
    J_k^+&=&(2k,2k+\alpha), \nonumber \\
    H_k^+&=&(2k+\alpha,2(k+1)), \nonumber \\
    J_k^-&=&(-2k-\alpha,-2k), \nonumber \\
    H_k^-&=&(-2(k+1),-2k-\alpha). \nonumber
    \end{eqnarray}
    Besides
    \begin{eqnarray}
    \mathcal{A}&=&\{2k+\alpha \; ;\; k=0\ldots N+1\}, \nonumber \\
    \mathcal{B}&=&\{-2k-\alpha \; ;\; k=0\ldots N\}, \nonumber \\
    \mathcal{C}&=&\{2k \; ;\; k=-N\ldots N+1\}, \nonumber \\
    \mathcal{D}&=&\{2k+1 \; ;\; k=-N\ldots N\}. \nonumber
    \end{eqnarray}
    The period of any trajectory starting in $[-\xi,2+\xi]$
    is given in Table \ref{tab:T}, where we have defined an
    auxiliary quantity $\eta$ when appropriate.
  \end{enumerate}
\end{enumerate}

\begin{table}
\caption{\label{tab:T} Period of the trajectories according to the
initial point $z_0\in [-\xi,2+\xi]$, with $b=2$ and $\xi > 2$, in
equation (\ref{eq:zlog}).}
\begin{center}
\begin{tabular}{lccr}
\hline
Initial point & $\eta$ & Period & Condition\\
\hline \multirow{2}{*}{$z_0 \in J_k^\pm$ and $\notin \mathcal{D}$} &
\multirow{2}{*}{$-2k\pm z_0$} & $4N+6$ &\mbox{ $\eta+\alpha>2$}\\
 &  & $4N+4$&\mbox{$\eta+\alpha<2$}\\
\hline \multirow{2}{*}{$z_0 \in H_k^\pm$ and $\notin \mathcal{D}$} &
\multirow{2}{*}{$2(k+1) \mp z_0$} & $4N+2$ & \mbox{ $\eta>\alpha$}\\
 &  & $4N+4$ & \mbox{ $\eta<\alpha$} \\
\hline \multirow{2}{*}{$z_0 \in \mathcal{A \bigcup B}$} &
\multirow{2}{*}{} & $4N+6$ &\mbox{ $\alpha>1$}\\
 &  & $4N+4$&\mbox{$\alpha<1$} \\
\hline \multirow{2}{*}{$z_0 \in \mathcal{D}$} &
\multirow{2}{*}{} & $2N+3$ &\mbox{ $\alpha>1$}\\
 &  & $2N+1$&\mbox{$\alpha<1$} \\
\hline
 $z_0 \in \mathcal{C}$ & & $2N+2$& $\forall \alpha$\\
\hline
\end{tabular}
\end{center}
\end{table}


\setcounter{section}{0}

\clearpage
\clearpage
\clearpage
\clearpage
\clearpage
\clearpage

\bigskip
\bigskip

\begin{thebibliography}{99}

\bibitem{May2} May R M 1976  Simple mathematical-models with very complicated dynamics
\emph{Nature }\textbf{261} 459--67

\bibitem{mayetal} May R M, Conway G R, Hassell M P and Southwood T R E 1974
Time delays, density-dependence and single-species oscillations
\emph{J. Anim. Ecol.} \textbf{43} 747--70

\bibitem{hassell} Hassell M P 1975 Density-dependence in single-species populations
\emph{J. Anim. Ecol.} \textbf{44} 283--95

\bibitem{Varley} Varley G C, Gradwell G R and Hassell M P 1973
\emph{Insect population ecology: an analytical approach}
(Oxford: Blackwell)

\bibitem{May1} May R M and Oster G F 1976 Bifurcations and dynamic complexity
in simple ecological models
\emph{Am. Nat.} \textbf{110} 573-99

\bibitem{May3} May R M 1975 Biological populations obeying difference equations:
Stable points, stable cycles, and chaos
\emph{J. theor. Biol.} \textbf{51}  511--24

\bibitem{power} Banerjee S and Verfghese G C 2001 \emph{Nonlinear
phenomena in power electronics: Attractors, bifurcations, chaos and
nonlinear control} (New York: IEEE Press)

\bibitem{Hogan} Hogan S J, Higham L  and Griffin T C L 2007
Dynamics of a piecewise linear map with a gap \emph{Proc. R. Soc. A}
\textbf{463} 49--65

\bibitem{Hilborn} Hilborn R C 2000 \emph{Chaos and nonlinear dynamics}
(Oxford University Press)

\bibitem{Strogatz} Strogatz S H 1994 \textsl{Nonlinear Dynamics and
Chaos} (Cambridge MA: Perseus Books)

\bibitem{Collet} Collet P and Eckmann J P 1980 \emph{Iterated Maps on the
Interval as Dynamical Systems} (Birkh\"auser)

\bibitem{Ott} Ott E 2002 \emph{Chaos in dynamical systems}
(Cambridge University Press)

\bibitem{OR} Oteo J A and Ros J 2007
Double precision errors in the logistic map: Statistical
study and dynamical interpretation
\emph{Phys. Rev. E} {\bf 76} 036214--22

\bibitem{Bailey} Bailey D H 1993
Algorithm 719: Multiprecision translation and execution of Fortran programs
\emph{ACM Trans. Math. Software }{\bf 19} 288--319

\bibitem{Bailey1} Bailey D H 2005
High-precision floating-point arithmetic in scientific computation
\emph{Comput. Sci. Eng.} {\bf 7} 54--61

\bibitem{Costa} Costa U M S, Lyra M L, Plastino A R and Tsallis C
1997 Power-law sensitivity to initial conditions within a
logisticlike family of maps: Fractality and nonextensivity
\emph{Phys. Rev. E} {\bf 56} 245--50

\bibitem{Tsallis} Tsallis C, Plastino A R and Zheng W M
Power-law sensitivity to initial conditiions: New entropic
representation 1997 \emph{Chaos, Solitons \& Fractals} {\bf 8}
885--91

\bibitem{Grebogi-Ott} Grebogi C, Ott E and Yorke J A 1983
Crises, sudden changes in chaotic attractors, and transient chaos
\emph{Physca D} {\bf 7} 181--200

\bibitem{Jensen} Jensen R V and Myers C R 1985
Images of the critical points of nonlinear maps
\emph{Phys. Rev. A} {\bf 32} 1222--4

\bibitem{Eidson} Eidson J, Flynn S, Holm C, Weeks D and Fox R F 1986
Elementary explanation of boundary shading in chaotic-attractor plots
for the Feigenbaum map and the circle map
\emph{Phys. Rev. A} {\bf 33} 2809--12

\bibitem{Feigen1} Feigenbaum M J 1978 Quantitative universality for a class of
non-linear transformations
\emph{J. Stat. Phys.} {\bf 19} 25--52

\bibitem{Feigen2} Feigenbaum M J 1979 Universal metric properties of
non-linear transformations
\emph{J. Stat. Phys.} {\bf 21} 669--706

\bibitem{Lai} Lai Y C, Grebogi C and Yorke J A 1992 Sudden change in
the size of chaotic attractors: How does it occur? \emph{Applied
Chaos} (chapter 19) Ed. by Kim H and Stringer J (John Wiley \& Sons,
Inc.)

\bibitem{Schlogl} Beck C and Schl\"ogl F 1993 \emph{Thermodinamics of chaotic
systems} (Cambridge University Press)

\bibitem{Maistrenko} Maistrenko Yu L, Maistrenko V L and
Chua L O 1993 Cycles in chaotic intervals in a time-delayed Chua's
circuit \emph{Int. J. Bifurcation and Chaos} \textbf{3} 1557--72

\bibitem{Romeiras} Grebogi C, Ott E, Romeiras F and Yorke J A 1987
Critical exponents for crisis-induced intermittency \emph{Phys Rev
E} {\bf 36} 5365--79

\bibitem{Diamond} Diamond P, Kloeden P, Pokrovskii A and Vladimirov
A 1995 Collapsing effects in numerical simulations of a class of
chaotic dynamical systems and random mappings with a single
attracting center \emph{Physica D} {\bf 86} 559--71

\bibitem{Yuan} Yuan G and Yorke J A 2000
Collapsing of chaos in one dimensional maps \emph{Physica D} {\bf
136} 18--30

\bibitem{Banerjee1} Banerjee S, Karthik M S,  Yuan G and Yorke J A 2000
Bifurcations in one-dimensional piecewise smoth maps: Theory and applications
in switching circuits
\emph{IEEE Trans. Circuits Syst.-I: Fund. Theory Appl. }\textbf{47} 389--94

\bibitem{Avrutin3} Avrutin V and Schanz M 2006
On multi-parametric bifurcations in a scalar piecewise linear map
\emph{Nonlinearity} \textbf{19} 531--52

\bibitem{Mehra} Mehra V and Ramaswamy R 1996
Maximal Lyapunov exponent at crises \emph{Phys. Rev. E} {\bf 53}
3420-4

\end{thebibliography}
\end{document}